\documentstyle{mn}
\newif\ifAMStwofonts
\def\cf{{\rm cf.}}
\def\eg{{\rm e.g.}}

\def\etal{{\it et al.}}

\ifoldfss
  \ifCUPmtlplainloaded \else
    \NewTextAlphabet{textbfit} {cmbxti10} {}
    \NewTextAlphabet{textbfss} {cmssbx10} {}
    \NewMathAlphabet{mathbfit} {cmbxti10} {} 
    \NewMathAlphabet{mathbfss} {cmssbx10} {} 
  \fi
  \ifAMStwofonts
    \ifCUPmtlplainloaded \else
      \NewSymbolFont{upmath} {eurm10}
      \NewSymbolFont{AMSa} {msam10}
      \NewMathSymbol{\upi}     {0}{upmath}{19}
      \NewMathSymbol{\umu}     {0}{upmath}{16}
      \NewMathSymbol{\upartial}{0}{upmath}{40}
      \NewMathSymbol{\leqslant}{3}{AMSa}{36}
      \NewMathSymbol{\geqslant}{3}{AMSa}{3E}

      \let\leq=\leqslant 
      \let\geq=\geqslant \let\ge=\geqslant
    \fi
  \fi
\fi 

\ifnfssone
  \newmathalphabet{\mathit}
  \addtoversion{normal}{\mathit}{cmr}{m}{it}
  \addtoversion{bold}{\mathit}{cmr}{bx}{it}
  \newmathalphabet{\mathbfit} 
  \addtoversion{normal}{\mathbfit}{cmr}{bx}{it}
  \addtoversion{bold}{\mathbfit}{cmr}{bx}{it}
  \newmathalphabet{\mathbfss} 
  \addtoversion{normal}{\mathbfss}{cmss}{bx}{n}
  \addtoversion{bold}{\mathbfss}{cmss}{bx}{n}
  \ifAMStwofonts
    \ifCUPmtlplainloaded \else
      \UseAMStwoboldmath
      \makeatletter
      \new@mathgroup\upmath@group
      \define@mathgroup\mv@normal\upmath@group{eur}{m}{n}
      \define@mathgroup\mv@bold\upmath@group{eur}{b}{n}
      \edef\UPM{\hexnumber\upmath@group}
      \new@mathgroup\amsa@group
      \define@mathgroup\mv@normal\amsa@group{msa}{m}{n}
      \define@mathgroup\mv@bold\amsa@group{msa}{m}{n}
      \edef\AMSa{\hexnumber\amsa@group}
      \makeatother
      \mathchardef\upi="0\UPM19
      \mathchardef\umu="0\UPM16
      \mathchardef\upartial="0\UPM40
      \mathchardef\leqslant="3\AMSa36
      \mathchardef\geqslant="3\AMSa3E

      \let\leq=\leqslant 
      \let\geq=\geqslant \let\ge=\geqslant
    \fi
  \fi
\fi 

\ifnfsstwo
  \DeclareMathAlphabet{\mathbfit}{OT1}{cmr}{bx}{it}
  \SetMathAlphabet\mathbfit{bold}{OT1}{cmr}{bx}{it}
  \DeclareMathAlphabet{\mathbfss}{OT1}{cmss}{bx}{n}
  \SetMathAlphabet\mathbfss{bold}{OT1}{cmss}{bx}{n}
  \ifAMStwofonts
    \ifCUPmtlplainloaded \else
      \DeclareSymbolFont{UPM}{U}{eur}{m}{n}
      \SetSymbolFont{UPM}{bold}{U}{eur}{b}{n}
      \DeclareSymbolFont{AMSa}{U}{msa}{m}{n}
      \DeclareMathSymbol{\upi}{0}{UPM}{"19}
      \DeclareMathSymbol{\umu}{0}{UPM}{"16}
      \DeclareMathSymbol{\upartial}{0}{UPM}{"40}
      \DeclareMathSymbol{\leqslant}{3}{AMSa}{"36}
      \DeclareMathSymbol{\geqslant}{3}{AMSa}{"3E}

      \let\leq=\leqslant 
      \let\geq=\geqslant \let\ge=\geqslant
    \fi
  \fi
\fi 

\ifCUPmtlplainloaded \else
  \ifAMStwofonts \else 
    \def\upi{\pi}
    \def\umu{\mu}
    \def\upartial{\partial}
  \fi
\fi

\title{Will GRB 990123 Perform an Encore?}
\author[Roger. D. Blandford  and David J. Helfand]
{Roger D. Blandford
\thanks{E-mail: rdb@tapir.caltech.edu.
Also at: Theoretical Astrophysics, 130-33 Caltech, Pasadena, CA 91125, USA} 
and 
David J. Helfand
\thanks{E-mail: djh@astro.ca.ac.uk.
Also at: Dept. Astronomy, Columbia University,
          538 W 120th St, New York, NY10027, USA}\\
Institute of Astronomy, Madingley Rd., Cambridge CB3 0HA}
\date{Accepted 1999.      Received 1999}
\pagerange{\pageref{firstpage}--\pageref{lastpage}}
\pubyear{1999}
\begin{document}
\maketitle
\label{firstpage}
\label{lastpage}
\begin{abstract}
The recent gamma ray burst, GRB 990123, has an absorption redshift $z_s=1.60$,
implying an apparent energy
$E\ge3\times10^{54}$~erg, and a peak luminosity $L_{{\rm max}}\ge6\times10^{53}$~erg s$^{-1}$, assuming isotropic emission.  This energy is ten times larger than
hitherto measured and in excess of the rest mass of the sun.  
Optical observations have revealed an associated galaxy
displaced from the line of sight by $\sim 0.6^{\prime\prime}$. This raises
the possibility that the burst is enhanced by gravitational lensing.
We argue that existing observations probably only allow magnifications $\mu>400$ 
if the galaxy is at $z_d=1.60$ and the burst originates at much higher 
redshift. It should
be possible to exclude this possibility by examining the burst time structure.
If, as we anticipate, multiple imaging can be excluded, GRB 990123 remains the 
most intrinsically luminous event yet observed in its entirety.  
\end{abstract}
\section{Introduction}
The majority of gamma ray bursts appear to be located at cosmological distances.
This raises the possibility that a small minority may be brightened anomalously
through being multiply imaged by an intervening galaxy.  This may lead to the detection
of multiple bursts (\eg, Paczy\'nski 1986), although the {\it a priori} probability of such an 
occurence is not high (\eg, Blaes \& Webster 1992). In the event of such a propitious alignment,
we stand to learn much about the source and the deflector.

The recent burst GRB 990123 \cite{piro} is the brightest yet detected by the
BeppoSAX Wide Field Camera; given the lower limit to its distance set by the detection of absorption lines in the spectrum of the optical transient (OT) at
$z_s=1.60$ \cite{kel99}, it is also the most luminous burst with a firm
distance limit: adopting a world model with
$h=0.6,\Omega_0=0.3,\Omega_\Lambda=0.7$, the observed 20-700 keV
X,$\gamma$-ray fluence of $3.5\times10^{-4}$~erg cm$^{-2}$ translates into a
minimum burst energy of $\sim3\times10^{54}{\cal B}$~erg, where ${\cal B}$ 
is the beaming fraction. This is ten times larger than hitherto reported
and, if the emission is isotropic and unmagnified, represents an energy in excess
of the rest mass of a neutron star, comprehensively ruling out many theoretical
models. In addition, the peak luminosity
during the burst can be estimated as $6\times10^{53}{\cal B}$~erg s$^{-1}$
which is $2\times10^{-6}{\cal B}c^5/G$. The reported optical emission is much smaller,
$\sim1.5\times10^{51}{\cal B}$~erg, though still in excess of the energy associated with a conventional supernova.

Although initial reports of a foreground galaxy \cite {ode99} within
$2^{\prime\prime}$ of the burst at a redshift $z\sim 0.29$
\cite{hjo99a}
have been discounted (Yadigaorglu {\it et al.}
1999; Hjorth {\it et al.} 1999b; Djorgovski {\it et al.} 1999a), the recent
HST image \cite{fru99} reveals a fainter galaxy centered only $\sim0.6^{\prime
\prime}$ from the burst afterglow. It is thus still
important to explore the possibility
that the burst has been magnified by lensing.  In this note, we examine, in more
detail, the possibility that lensing may be occurring and, if so, some of its
ramifications.

\section{Macrolensing}
The largest magnifications observed in known galaxy lenses are found in ``quad''
geometries associated with elliptical mass perturbations, when the source is located 
close to a caustic surface and two images straddle the critical curve.
In this case, the two bright bursts will be closely spaced in time and on the sky, 
and will be the  second and third to arrive. In addition to the fainter and
more widely separated events, labelled 1 and 4, there will also be a faint (or invisible) 
fifth image, located near the lens galaxy nucleus, which we shall ignore.

We make an elementary model of a nearly circular lens (\cf, Blandford \& Kovner 1988;
Schneider, Ehlers \& Falco 1992; Blandford \& Hogg 1996).
Let the scaled surface potential in the vicinity of the Einstein ring be written
\begin{equation}
\psi(\vec r)=f(r)+g(r)\cos2\phi,
\end{equation}
where $r$ is measured in units of the unperturbed Einstein ring radius, so that
$f'(1)=1$, and $g(r)$ is a perturbation which measures the
ellipticity in the potential and its radial 
variation.  As $f,g$ depend quite heavily upon the dark matter halo, the ellipticity can only be guessed, though the position angle probably agrees with that of the luminous matter. (Observed lenses often require external shear to
fit their image geometries.)

The time function is given by 
\begin{equation}
t=r^2/2-\psi-\vec r\cdot\vec\beta
\end{equation}
Images at $\vec r$ have sources at $\vec\beta$ located as extrema of $t$.
Hence, 
\begin{equation}
\vec\beta=[(1-f'')\delta-g'\cos2\phi]\vec{\hat r}+2g\sin2\phi\vec{\hat\phi}
\end{equation}
where $\delta=r-1$ and all derivatives are evaluated at $r=1$.
The Hessian matrix of $t$ is given by 
\begin{eqnarray}
H&=&\pmatrix{t_{,rr}&r^{-1}t_{,r\phi}-r^{-2}t_{,\phi}\cr 
r^{-1}t_{,r\phi}-r^{-2}t_{,\phi}&r^{-1}t_{,r}+r^{-2}t_{,\phi\phi}\cr}\cr
&=&\pmatrix{1-f''&2(g'-g)\sin2\phi\cr 
2(g'-g)\sin2\phi&(1-f'')\delta +(4g-g')\cos2\phi\cr}
\end{eqnarray}
in polar coordinates, (to lowest order), 
and the scalar magnification, $\mu$, is the reciprocal of its determinant.
Now, $\mu\rightarrow\infty$ when the image lies on the critical curve
\begin{equation}
\delta=\delta_c={(g'-4g)\cos2\phi\over1-f''}
\end{equation}
or equivalently, when the source lies on the caustic
\begin{eqnarray}
\beta=\beta_c&=&2g(-2\cos2\phi\vec{\hat r}+\sin2\phi\vec{\hat\phi})\cr
&=&4g(-\cos^3\phi\vec{\hat x}+\sin^3\phi\vec{\hat y}).
\end{eqnarray}

If we now displace the source perpendicular to the caustic,
a pair of images will separate in opposite directions from the critical curve
along a line with 
$\delta r=2(g-g')\sin2\phi\delta\phi/(1-f'')$ where $\delta\phi$
(assumed to be $<<1$) is the displacement
of either image from the critical curve.  Perturbing the Hessian, we
find that 
\begin{equation}
\mu_{2,3}^{-1}=|6(1-f'')g\sin2\phi_2\delta\phi|,
\end{equation}
for each of the neighbouring, bright images. 

Expanding the time delay to third order, we find that the time delay between the two 
bright bursts is given by
\begin{eqnarray}
t_3-t_2&=&4g\sin2\phi_2\delta\phi^3\cr
&=&{1\over54(1-f'')^3g^2\sin^22\phi_2\mu_{2,3}^3}
\end{eqnarray}
to leading order. These expressions must be modified near a cusp 
when $|\sin\phi_2|<|\mu_{2,3}|\Delta t$.
(Higher order catastrophes are possible, but less probable: \eg, Schneider \etal
 1992.)

We can also locate the preceding (1) and following (4) bursts in the high-magnification, 
small-ellipticity limit at position angles
\begin{eqnarray}
\sin\phi_{1}&=&[\sin\phi_2\{\cos^2\phi_2-(1-{1\over4}\sin^22\phi_2)^{1/2}\}]\cr
\sin\phi_{4}&=&[\sin\phi_2\{\cos^2\phi_2+(1-{1\over4}\sin^22\phi_2)^{1/2}\}]
\end{eqnarray}
Measuring $\phi_2$ from the minor axis, we find, without loss of generality, that 
when the merging pair is in the first quadrant,
the preceding burst is in the second quadrant and the following burst is in the fourth
quadrant. The corresponding delays are given by
\begin{eqnarray}
t_2-t_1&=&{(33\cos2\phi_2+2^{1/2}(7+\cos4\phi_2)^{3/2}-\cos6\phi_2)g\over8}\cr
t_4-t_2&=&{(-33\cos2\phi_2+2^{1/2}(7+\cos4\phi_2)^{3/2}+\cos6\phi_2)g\over8}\cr
&&
\end{eqnarray}
$t_2-t_1$, $(t_4-t_2)$ varies between $8g$, (0) and 0, ($8g$) as 
$\phi$ increases from 0 to $\pi/2$. The associated magnifications are given by
\begin{eqnarray}
\mu_1^{-1}={1\over4}[15\cos2\phi_2&+&2^{1/2}(5-\cos4\phi_2)(7+\cos4\phi_2)^{1/2}\cr
&+&\cos6\phi_2](1-f'')g\cr
\mu_4^{-1}={1\over4}[15\cos2\phi_2&-&2^{1/2}(5-\cos4\phi_2)(7+\cos4\phi_2)^{1/2}\cr
&+&\cos6\phi_2](1-f'')g\cr
&&
\end{eqnarray}
Thus, observation of either burst 2 or 3 allows one to specify completely the
magnifications, time intervals, and locations of the other three images.
These expressions are only valid as long as $\delta\phi<<1$ and the magnification is large, specifically as long as $\mu>>(6(1-f'')g)^{-1}$. 

When the source is even farther from the caustic, it is possible to create four,
similarly-magnified bursts. These will be located at the solutions of the
quartic
\begin{equation}
\beta_y\cos\phi-\beta_x\sin\phi=4g\sin\phi\cos\phi
\end{equation}
In this case, it is necessary to observe two bursts optically in order to solve
for the source location, $\vec\beta$.
The associated magnifications are given by
\begin{equation}
\mu^{-1}=|(4g\cos2\phi+\beta_x\cos\phi+\beta_y\cos\phi)(1-f'')|
\end{equation}
and the arrival times (ignoring a constant) by
\begin{equation}
t=-g\cos2\phi-\beta_x\cos\phi-\beta_y\sin\phi
\end{equation}
Note that if $\beta_x^{2/3}+\beta_y^{2/3}>(4g)^{2/3}$, then the source
is located outside the caustic and only two bursts will be seen.
Interestingly, if the source is located just outside of the cusp,
one of these bursts can be arbitrarily magnified and followed by a single, fainter
burst. However, this is a relatively rare occurence. Even less likely is a radial merger
geometry, when two, bright bursts, located much closer to the
galaxy nucleus, will follow an isolated burst.
Finally, if there is no multiple imaging, then the single burst will still be magnified
by a factor that depends upon the detailed mass distribution closer to the nucleus.
This factor is less than two for an isothermal sphere, as is typically assumed, and can only be large if 
the surface density is roughly constant at the observed image location. 

To summarize, if we are able to locate a burst with respect to the deflector
galaxy and can guess the ellipticity of the potential, then, on the
hypothesis that the observed burst is the first of a merging pair (and the 
second overall), we have outlined a procedure for predicting the location, the magnification and the delay of the first, third, and fourth bursts.  
Multiple bursts can still occur without strong magnification, but in this case
we must observe another burst to make more predictions. 

\section{Microlensing and Millilensing}
If the lensing galaxy comprises mainly stars, then the optical depth in the vicinity
of the critical curve is automatically $\Sigma/\Sigma_{{\rm crit}}\sim0.5$.  This means that
microlensing variations are unavoidable if the source is sufficiently small. 
As the characteristic time delays associated with individual stars are 
$<100\mu$s, the arrival times, locations, and spectra of individual bursts 
should not be seriously affected.  However, significant
magnification fluctuations are possible as long as the source is smaller than 
$\sim10^{16}\mu^{-1/2}$~cm, which can be true when the burst is less than a day old.

The mass distribution of the  deflector galaxy is likely
to have additional perturbations associated with arms, bars, etc., especially if it is a spiral; indeed, this is commonly observed in galaxy lenses which do not
obey magnification scalings close to catastrophes.  This is known as
millilensing. If the time delay between two neighbouring bursts is $\Delta t$,
a perturbing mass as small as $\sim10^5(\Delta t/1{\rm s})$~M$_\odot$, in the
vicinity of the images, suffices to change the magnifications by $O(1)$.

\section{Application to GRB 990123}

Assume first that the galaxy observed in the HST image is at $z=0.29$.
(Although spectra of the OT have not confirmed reported absorption lines at
this redshift, the brightest galaxy in the vicinity does have
$z=0.28$ \cite{hjo99b}, and so it is not excluded that the faint galaxy closest to
the burst lies at this distance.) 
Adopting a magnitude of $V=24.4$ and a mass-to-light ratio 
at the rest effective frequency of $\nu_0=7\times10^{14}$~Hz ($\equiv$~B),
the luminosity is $\nu L_\nu(\nu_0)=3\times10^8$~L$_\odot\sim0.02 L^\ast$
ignoring the effects of reddening.  The critical surface density is
$\Sigma_{{\rm crit}}=0.45$~g cm$^{-2}$ and the requisite mass-to-light ratio
becomes $(M/L)_B=200(\theta_E/0.6'')^2$,
in solar units, for our world model, and where we have assumed
that all of the light is produced within the Einstein ring.  This is 
far too large for a galaxy at this redshift to produce multiple images.

However, it is also possible that the galaxy is at the absorption redshift
$z_d=1.6$ (Hjorth {\it et al.} 1999a; Djorgovski {\it et al.} 1999b) while the burst occured at a larger redshift. 
In this case, the K-band magnitude $K=21.6$
\cite{fru99} can be used to interpolate a rest B luminosity
$(\nu L_\nu)_B\sim 1.5\times10^{10}L_\odot\sim0.8L^\ast$. The galaxy is quite blue
suggesting that the reddening is not very large. 
For illustration, let us suppose that $z_s=3$ (our results are not very sensitive
to this choice). The critical density is $\Sigma_{{\rm crit}}=0.45$~g cm$^{-2}$ and
the required mass-to-light ratio has a value, $(M/L)_B\sim20$ in solar units if 
$\theta_E\sim0.6''$, (6~kpc), which, although large, cannot be ruled out. 
The observed galaxy has an axis ratio of $\sim4$.  However, the 
total mass distribution is likely to be more circular. We adopt, for illustration,
$f(r)=r,g(r)=0.1r$, consistent with a density axis ratio of $\sim2$, and we
measure $\phi=65^\circ$ for the observed burst. (This excludes a single, 
magnified, cusp image.) 

If we suppose that the observed burst was a merging double, then the first
and fourth bursts are located at $\phi_1 = 138^{\circ}$ and $\phi_4 = 273 ^{\circ}$.
(More detailed lens models 
do not change our qualitative conclusions and only affect them significantly, quantitatively,
through the scaling with $g$.) We can use the unit of time delay, which is 188~d, to compute 
the intervals:
\begin{eqnarray}
t_2-t_1&=&8 (g/0.1)~{\rm d}\cr
t_3-t_2&=&50\left({\mu_{2,3}\over100}\right)^{-3} (g/0.1)^{-2}~{\rm s}\cr
t_4-t_3&=&110 (g/0.1)~{\rm d}
\end{eqnarray}

Given the potentially short time interval $t_3-t_2$ for cases of large magnification, it is necessary 
to ask if the multiple bursts might have occured
within the 100-s duration of the burst itself. Cursory examination of the BATSE
light curve at 0.5~s resolution
indeed reveals two distinct peaks separated by $\sim 12$ seconds;
the obvious $\sim 25\%$ difference in peak flux could conceivably be produced
by differential milli- or microlensing along the two paths. However, examination of
the spectra of the two peaks shows that the second is distinctly softer:
taking 8-s intervals (approximately the FWHM) centered on each peak yields
count ratios of 1.07, 1.09, 1.11, 1.41, and 1.82 in the 20-50 keV, 50-100 keV,
100-300 keV, 200-1600 keV, and 600-11,000 keV bands respectively, where the
first three data points are derived from BATSE \cite{kip99}
and the last two from COMPTEL \cite{con99}. In addition, the distinct
peak 76 seconds after the BATSE trigger
has no counterpart with the same separation as the earlier pair of peaks; the
closest local maximum is over 19 seconds away. Finally, the overall spectral
evolution of the burst is from hard to soft as is typical for BATSE bursts
(\eg, Preece {\it et al.} 1998). We conclude that the 100-second long burst profile does not
conceal a temporally resolved double burst at $\sim1$~s resolution. 

The lower limit on this interval can be extended to several tens of minutes depending
on the location of the observing satellites with respect to the
Earth and the South Atlantic Anomaly (SAA) at the time of the burst. SAX,
for example, saw no burst within a factor of 40 in intensity in the Wide Field
Camera from this location for the 1450 s preceeding the burst (following the
satellite's emergence from the SAA);
after the burst, only 170~s elapsed before the Earth occulted the burst
position (SAX Team, private communication). In BATSE, the burst was observed
$64^\circ$ above the horizon, implying that the source remained visible
for at least $\sim 15$ min after the trigger; thus, $\pm900$~s is a
conservative lower limit for the interval between two resolved bursts. (Also relevant are the data from
Ulysses which saw no burst consistent with the 
location of GRB 990123 for a period of at least three days before and after the event (K. Hurley, private
communication), although coverage was only about 80\% complete and we
cannot completely exclude a second burst.)

If, as we argue, $1~s\leq t_3-t_2 \leq 900~s$ is excluded, then so are
magnifications $40\leq \mu_{2,3} \leq 400$. Furthermore, we can use
limits on additional point sources in the HST image within
$2^{\prime\prime}$ of the afterglow to place constraints on
lensed images. If $\mu_{2,3}<40$, then using
$\mu_1=4$ (from eq. 11), we find that burst 1 would have had a fluence $>3\times10^{-5}$~erg cm$^{-2}$
on or around Jan. 15. It is unlikely, though not completely excluded, that 
this failed to trigger any detector. However,
the afterglow associated with burst 1 would have been
brighter than $V\sim28.8$ at the time of the HST image even allowing for its
additional fading with time. We have performed
two-dimensional Gaussian fits (including
sloping, planar baselines) to all local maxima within $2^{\prime\prime}$ of
the OT in the HST image. The only feature consistent with the psf (derived from
a similar fit to the OT as FWHM = $3.2 \pm 0.1$ pixels) is the faint object located
$1.4^{\prime\prime}$ north of the OT. This $\sim 4 \sigma$ excess has a
magnitude of approximately $V \sim 28.4$ (scaled to the value of $V=25.2$ for
the OT reported by Bloom et al. (1999)). We take this as 
an upper limit to the magnitude of the first afterglow. We can therefore almost
exclude $\mu_{2,3} \leq 40$.

The remaining parameter space is described by $t_{2,3} \leq 1~s$ and
$\mu_{2,3} \geq 400$. In this case, the first afterglow will
be undetectable. However, the finite size of the source becomes a factor at
these high magnifications. For a source angular size B(t), the magnification
is limited to
\begin{equation}
\mu_{2,3}(t) < (6g{\rm sin}2\phi B(t)/\theta_{E})^{-1/2} \sim 400(B/8\mu
{\rm as})^{-1/2}.
\end{equation}
This limit should not affect the burst itself, although it will eventually
influence the afterglow. Unfortunately, our lack of understanding of the ambient
environment and the nature of the explosion precludes a confident expression for B(t). 
However, a naive estimate for a spherical, relativistic blastwave with
$E\sim10^{52}$~erg and $n\sim 1$~${\rm cm}^{-3}$ (e.g., Blandford and McKee 1977)
gives $B\sim2(t/1{\rm d})^{5/8} \mu {\rm as}$. This limits the magnification to
$\mu_{2,3}\leq 600(t/17{\rm d})^{-0.3}$. After this inequality is violated, the afterglow 
emission will decline correspondingly more steeply with time. In fact, just 
such an increase in the rate of decline has been reported at $t=11$~d
(Yadigaroglu \& Halpern 1999). We therefore
cannot confidently exclude lensing with
$\mu_{2,3}\sim400$ at this stage. However, if it is possible to examine
the subsecond time variations in the BATSE lightcurve of
GRB980123 and thereby limit $t_3-t_2$
to $\approx10$~ms, then $\mu_{2,3}$ would have to exceed $\approx1000$, and
all multiple imaging by a $z_d=1.6$ deflector would effectively be ruled out.

In summary, three arguments (the high $M/L$ of the galaxy, the implausibility of missing
the first burst and of failing to detect its 
afterglow) can already be marshalled against the lensing hypothesis.  Three additional
steps might effectively eliminate it - searching for double structure on 
subsecond timescales in the BATSE data,  setting a better limit on 
the presence of additional afterglow images at the predicted locations
and obtaining a reliable photometric or spectroscopic redshift for the galaxy.
Contrariwise, if it turns out that the burst was highly magnified
by lensing, then the burst energy 
would be reduced to $\sim9\times10^{51}{\cal B}(2\mu_{2,3}/1000)^{-1}$ erg. 

\section{Conclusion}
GRB 990123 serves as a reminder that 
multiple imaging of a gamma-ray burst is to be expected 
eventually in a large enough sample and the analysis of \S2 should be generally
applicable.
While we cannot completely rule out the possibility that it
has been multiply imaged and strongly magnified, it should be possible to do so
soon. In this case, if we have not observed (or do not observe) an echo of GRB 990123, then
the magnification is limited to $\mu\sim2$, except 
under quite contrived models, leaving GRB 990123 as the 
most intrinsically luminous cosmic event yet observed in its entirety.
\section*{Acknowledgements}
We thank Martin Rees for encouragement, the Institute of Astronomy, University of Cambridge for hospitality, and the
Beverly and Raymond Sackler Foundation for support during the
preparation of this paper. Support under NSF grant AST 95-29170 and NASA grant
5-2837 is also gratefully acknowledged.

\end{document}